%% file: ICCAD2024.tex
\documentclass[sigconf,screen=true,anonymous=false,bookmarks=false,nonacm]{acmart}
\input{setting-acm}

\graphicspath{{./figs/}}
\setcopyright{none}

\newcommand\eat[1]{}
 
\allowdisplaybreaks

\title{LiTformer: Efficient Modeling and Analysis of High-Speed Link Transmitters Using Non-Autoregressive Transformer}

\author{
 \small
    Songyu Sun$^1$, Xiao Dong$^1$, Yanliang Sha$^2$, Quan Chen$^2$, Cheng Zhuo$^{1,3,*}$ \\
 $^1$Zhejiang University, Hangzhou, China; $^2$Southern University of Science and Technology, Shenzhen, China\\$^3$Key Laboratory of Collaborative Sensing and Autonomous Unmanned Systems of Zhejiang Province, Hangzhou, China\\$^*$Corresponding email: czhuo@zju.edu.cn \\
}

\begin{document}
\input{docs/abstract}

\maketitle
\pagestyle{empty}
\input{docs/intro}
\input{docs/bkgd}
\input{docs/prelim}
\input{docs/formulation}
\input{docs/model}
\input{docs/traintest}
\input{docs/exp}

\input{ICCAD2024.bbl}

\bibliographystyle{IEEEtran}

\end{document}

%% file: setting-acm.tex
\iftrue
\usepackage{titlesec}
\titlespacing\section{2pt}{5pt plus 1pt minus 1pt}{0pt plus 1pt minus 1pt}
\titlespacing\subsection{2pt}{5pt plus 1pt minus 1pt}{0pt plus 1pt minus 1pt}
\titlespacing\subsubsection{2pt}{5pt plus 1pt minus 1pt}{2pt plus 1pt minus 1pt}
\fi

\iftrue
\fi
\usepackage{lipsum}
\usepackage{graphicx}
\usepackage{amsmath}
\usepackage{footnote}
\usepackage{mathtools}
\usepackage{mathrsfs}     % mathscr command
\usepackage{comment}
\usepackage{color}
\usepackage[subrefformat=parens,labelformat=parens]{subfig}
\usepackage{bm,bbm}
\usepackage{multirow}
\usepackage{threeparttable,booktabs}
\usepackage{blkarray}
\usepackage{tikz}
\usetikzlibrary{positioning,calc,fit,decorations.pathmorphing,shapes.geometric,shapes.gates.logic.US,calc}
\usepackage{balance}
\usepackage{courier}  
\usepackage{ellipsis}

\usepackage{cleveref}                                      % smart citation
\usepackage[mathcal]{eucal}
\usepackage[]{algpseudocode}                               % algorithm package
\algrenewcommand\textproc{\texttt}
\makeatletter\let\float@addtolists\relax\makeatother
\usepackage{algorithm}
\usepackage{filecontents}                                  % support to pgfplots
\usepackage{pgfplots}
\usepackage{pgfplotstable}
\usepackage{pgf-pie}
\usepgfplotslibrary{groupplots}
\pgfplotsset{compat=newest}
\usepackage[figuresright]{rotating}
\usepackage{xcolor,colortbl}   
\usepackage{amsmath}
\theoremstyle{plain}

\theoremstyle{definition}

\algrenewcommand\textproc{\texttt}

\usepackage[skip=1pt]{caption}            % set space between figure and caption
\definecolor{CUHKorange}{RGB}{244,106,18} %F47012
\definecolor{CUHKblue}{RGB}{0,111,190}    %006FBE
\definecolor{CUHKgreen}{RGB}{0,127,128}   %007F80
\definecolor{CUHKred}{RGB}{228,46,36}     %E42E24
\definecolor{CUHKyellow}{RGB}{198,148,34} %C69422
\definecolor{CUHKdark}{RGB}{114,44,114}   %722C72
\definecolor{CUHKmiddle}{RGB}{144,44,144} %902C90

\usepackage{tcolorbox}
\tcbuselibrary{skins,breakable}
    {\endtcolorbox}

%% file: docs/abstract.tex
\begin{abstract}
High-speed serial links are fundamental to energy-efficient and high-performance computing systems such as artificial intelligence, 5G mobile and automotive, enabling low-latency and high-bandwidth communication. Transmitters (TXs) within these links are key to signal quality, while their modeling presents challenges due to nonlinear behavior and dynamic interactions with links. In this paper, we propose LiTformer: a Transformer-based model for high-speed link TXs, with a non-sequential encoder and a Transformer decoder to incorporate link parameters and capture long-range dependencies of output signals. We employ a non-autoregressive mechanism in model training and inference for parallel prediction of the signal sequence. LiTformer achieves precise TX modeling considering link impacts including crosstalk from multiple links, and provides fast prediction for various long-sequence signals with high data rates. Experimental results show that LiTformer achieves 148-456$\times$ speedup for 2-link TXs and 404-944$\times$ speedup for 16-link with mean relative errors of 0.68-1.25\%, supporting 4-bit signals at Gbps data rates of single-ended and differential TXs, as well as PAM4 TXs.

\end{abstract}

%% file: docs/intro.tex
\section{Introduction}
The increasing demand for advanced computing capabilities in emerging data-driven applications, such as artificial intelligence (AI), 5G mobile networks, and automotive technologies, emphasizes the need for systems that are energy-efficient, cost-effective, and high-performance~\cite{muralidhar2022energy,shi2022toward,9874808,9856879}. In recent years, as the rising costs of silicon manufacturing and constraints in on-chip integration density continue to challenge the industry, chiplet-based high-density heterogeneous integration (HDHI) has emerged and is increasingly prevalent in various applications~\cite{10473908,ucie2022}.

\begin{figure}[t]
\centering
\includegraphics[width=\linewidth]{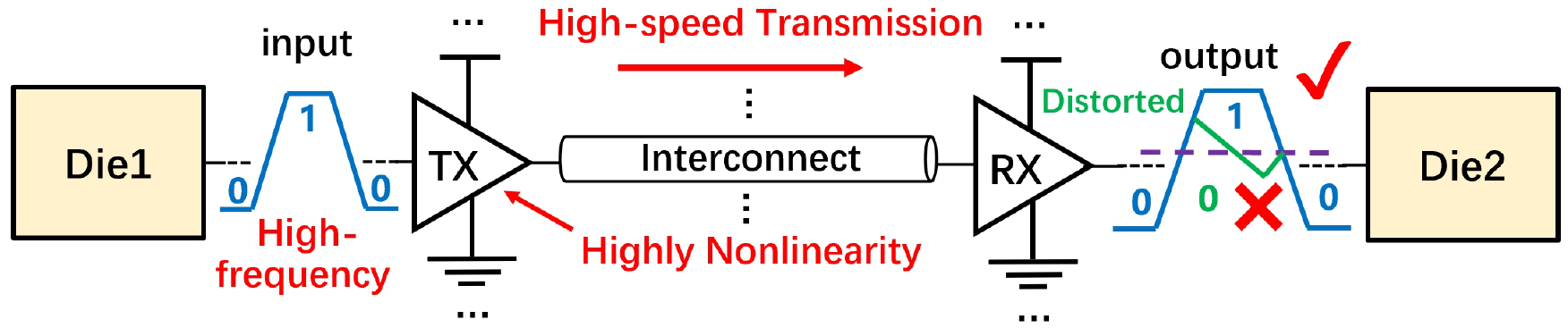}
\caption{An example of inter-chip high-speed link.}
\label{fig:link}
\end{figure}

High-speed serial links, essential for low-latency communication, rapid data transfer, and effective data processing, form the backbone of such advanced high-speed systems. These links typically include high-speed transmitters (TXs), interconnects (transmission lines), and receivers (RXs), as illustrated in~\Cref{fig:link}~\cite{chiu2019digital,ucie2022}. To accommodate the growing need for high-bandwidth and efficient communication, these links feature extensive density, with hundreds to thousands of signal pathways, and operate at high frequencies and data rates up to gigabits per second (Gbps), which has to address non-trivial signal integrity (SI) issues~\cite{ucie2022,li2020chiplet, 8811046}, such as crosstalk, signal attenuation, electromagnetic interference (EMI), $etc.$

In the high-speed links, TXs are one of the most crucial and resource-intensive components, as their performance directly affects the quality of the initial transmitted signals and the overall link integrity~\cite{1249467}. A degraded output signal from the TX can lead to significant distortion at the final RX, resulting in erroneous recovery of transmitted signals. To maintain high-quality and high-speed output signals, which are essential for preserving SI across the transmission path, TXs must operate at high frequencies and minimize timing errors due to process, voltage, temperature (PVT) variations, varying load conditions, $etc.$ They must also adhere to design metrics such as slew rate, signal swing, and equalization~\cite{5440927}. Thus, developing efficient models for these TXs is crucial for accurately predicting signal distortion and inter-symbol interference (ISI), thereby enhancing data transmission quality and reducing error rates. However, the evaluation of TXs in high-speed links often poses several challenges:
\begin{itemize}
\item High-speed TXs inherently possess strong nonlinear behaviour, which is further aggravated by parasitics especially when operating at the elevated Gbps data rates, significantly affecting the signal amplitudes or effective spectrum. Accurately capturing such robust nonlinearity at high frequencies is challenging and vital for TX modeling~\cite{663559}.
\item The transmission lines and RXs, as loads within the links, will significantly impact TX outputs~\cite{10196036}, in additional to complex crosstalk from dense links, which makes TX analysis very complicated and time-consuming. It is insufficient to simplify the problem of modeling TXs merely into modeling input-output signal characteristics of nonlinear circuits. It is hence critical to consider TX dynamics within the links, modeling the impacts of link parameters and crosstalk.
\item The nonlinearity and ``memory'' effect of TXs imply that an individual bit can affect several subsequent bits~\cite{jiao2017fast}. Multi-bit effects cannot be simply formulated using single-bit pulse response~\cite{hu2011comparative,kim2013eye} or edge response~\cite{shi2008efficient, park2017novel,chu2019statistical} via linear time-invariant (LTI) principles. Instead, it is necessary to directly predict multi-bit output sequences of TXs.
\end{itemize}

The most accurate way to analyze TX behavior in many links is using transistor-level models like SPICE models~\cite{synopsys}. However, they are very time-consuming due to the complex internal circuit details. Empirical behavioral models such as current source-based models (CSMs) and I/O buffer information specification (IBIS) models exhibit fast speed but usually have limited accuracy or capabilities in modeling complex interdependencies~\cite{ibis_ver7_2,1688797,4397344}.

Recently, Artificial Neural Networks (ANNs) have gained interest from both academia and industry, which significantly enhance the speed and efficiency of circuit modeling and simulation, especially for nonlinear components and systems~\cite{10196036,zhao2022modeling,NAGHIBI201966,1097995,Naghibi2019,Noohi2021ModelingAI,9395609}.
The inefficiency of static neural networks in modeling time-series sequences has further spurred research into time-domain sequence-to-sequence (seq2seq) neural networks for nonlinear circuit macromodeling~\cite{NAGHIBI201966,1097995,Naghibi2019,Noohi2021ModelingAI,9395609}, $e.g.$, Recurrent Neural Networks (RNNs) and Long Short-Term Memory networks (LSTMs)~\cite{9868892,NAGHIBI201966,Noohi2021ModelingAI, 9395609}. While these models can capture the internal dependency of temporal signal sequences through a recurrent structure, they struggle with vanishing/exploding gradients as well as extensive training time, making them unsuitable for highly nonlinear long-sequence signals with long-range dependencies. Their sequential nature also complicates the handling of non-sequential link parameters, limiting their effectiveness in modeling TX behavior considering various link parameters. In contrast, Y. Zhao $et al.$ developed a static Feedforward NN (FNN)-based model that includes link parameters but neglects the internal dependency of the signal sequence and overlooks crosstalk, only supporting weakly nonlinear signals from a single link TX~\cite{10196036,zhao2022modeling}.

For the future deployment of TX modeling in high-speed and high-density links, it is critical to: \begin{itemize}
    \item Effectively incorporate both input signals and link parameters into the model to capture TX dynamics within links; 
    \item Accurately capture long-range dependencies within the signal for precise long-sequence signal prediction;
    \item Efficiently account for complex crosstalk from multiple links.
\end{itemize} In this paper, to tackle the aforementioned challenges, we propose \textbf{LiTformer}, \textit{a non-autoregressive Trans\textbf{former}-based model for high-speed \textbf{li}nk \textbf{T}Xs}, utilizing a non-sequence-to-sequence (nonseq2seq) encoder-decoder architecture. This model employs a non-sequential encoder and a Transformer decoder to achieve precise TX modeling considering link impacts, including crosstalk from multiple links, and provides fast predictions for various long-sequence signals with high data rates. To the best of our knowledge, this is a pioneering effort to apply Transformer-based models for macromodeling nonlinear circuits. Our main contributions are listed below:
\begin{itemize}
    \item We propose \textbf{a novel Transformer-based model with a nonseq2seq encoder-decoder architecture for nonlinear TX modeling}, which not only considers both input signals and link characteristics including crosstalk, but also effectively captures long-range dependencies within the output signal sequences through the attention mechanism.
   \item We utilize \textbf{a non-autoregressive (NAR) approach for model training and inference}. We introduce an innovative one-pass filtered decoding technique that combines a single instance of parallel decoding with a single filtering step, ensuring short inference time and sufficient accuracy for long-sequence signals.
    \item The proposed \textit{LiTformer} can \textbf{efficiently predict arbitrary 4-bit output signals at Gbps data rates for TXs across various links}, supporting single-ended and differential TXs, as well as \textit{PAM4} (4-Level Pulse Amplitude Modulation) TXs.
\end{itemize}

Experimental results show LiTformer's efficiency in modeling high-speed link TXs for highly nonlinear long-sequence signals, accounting for different link parameters and crosstalk in many links. Our LiTformer demonstrates superior accuracy over previous works in modeling nonlinear TXs~\cite{10196036,9395609}. When compared to SPICE~\cite{synopsys}, the proposed LiTformer achieves up to 148-456$\times$ speedup for 2-link TXs and 404-944$\times$ speedup for 16-link TXs with minimal deviations between 0.68-1.25\%, supporting 4-bit signals with Gbps data rates of single-ended and differential TXs, as well as PAM4 TXs.

%% file: docs/bkgd.tex
\section{Background}
\subsection{Link Impact on TX Performance}
\label{subsec:link-effects}
In a set of links, the TX output is significantly impacted by its loading transmission lines and RXs via reflections, absorption, $etc.$ These effects also cause near end crosstalk (NEXT) and signal distortions when undesired coupling from adjacent lines interferes with the TX signal, which is more complex in high-speed dense links introducing timing errors and data corruption. In practice, transmission lines and RXs vary greatly to adapt to various situations and ever-evolving demands of communication technologies. As a result, there is a critical need to consider the dynamic interaction between TXs and links, which necessitates TX models to effectively capture the varying characteristics of different links.

\subsection{Characterization of Multi-Bit Response}
Due to their nonlinear nature, most TXs exhibit ``memory'' effects where an output bit affects its subsequent bits, causing a ripple effect down the sequence. For a TX with $m$-bit memory, the response to the current bit $bit(0)$ dependent on the past $m-1$ bits can be expressed as a nonlinear function of time $t$ and $m$ bits $y(bit(-m+1), \ldots, bit(-1), bit(0), t)$. The reactions within the highly nonlinear output signal of high-speed TXs have long-range dependencies. To evaluate overall performance, it is crucial to determine all $2^m$ responses to $m$-bit random inputs for TXs~\cite{jiao2017fast}. 

\begin{figure}[tb]
\centering
\includegraphics[width=\linewidth]{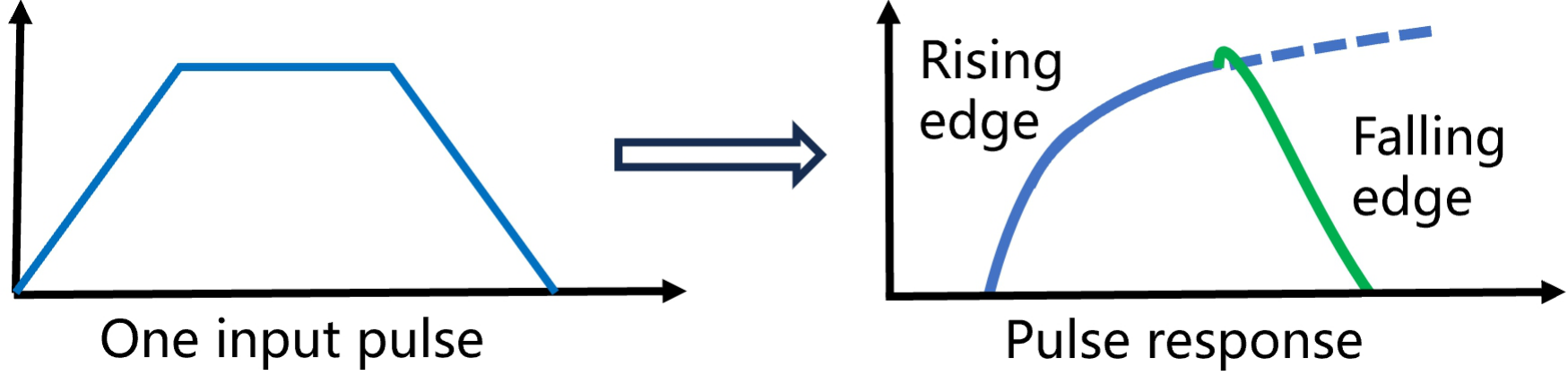}
\caption{Pulse response of the highly nonlinear TX.}
\label{fig:sbr-method}
\end{figure}

The asymmetry between the rising and falling edges of the pulse response prevents synthesizing the $2^m$ responses by time-shifting and superposing unit pulses or single bit responses (SBR) via LTI principles~\cite{hu2011comparative,kim2013eye}. Moreover, with increased nonlinearity at high frequencies, the TX output may drop before rising to stable as shown in~\Cref{fig:sbr-method}. Hence, methods that superpose edge responses like double-edge responses (DER) and multiple edge responses (MER)~\cite{shi2008efficient, park2017novel,chu2019statistical} fail to completely capture the output. Accurate TX output representation necessitates a direct characterization of the entire multi-bit signal.

\subsection{Deep Learning Model for Nonlinear TX}
\subsubsection{Related Works.}
ANN-based models have been widely studied for modeling nonlinear circuits in recent years~\cite{10196036,zhao2022modeling,NAGHIBI201966,1097995,Naghibi2019,Noohi2021ModelingAI,9395609}. 
Characterizing the transient input-output behaviour of nonlinear circuits can be treated as a seq2seq task, leading researchers to apply time-domain recurrent neural networks, $e.g.$, dynamic neural networks (DNNs)~\cite{1097995}, time-delay neural networks (TDNNs)~\cite{Naghibi2019}, RNNs and LSTMs~\cite{NAGHIBI201966,Noohi2021ModelingAI, 9395609}, to capture sequence dependencies. However, these models could not learn long-range dependencies in highly nonlinear signals due to vanishing/exploding gradients. Their time-sequential nature also renders them non-parallelizable, leading to extensive training and inference times. Moreover, they struggle with handling non-sequential link parameters for accurate modeling of TXs in the links. As a result, these seq2seq models are inadequate for nonlinear TX modeling. Only a few studies, like FNNs in~\cite{10196036,zhao2022modeling}, consider link parameters in TX modeling. However, FNNs assume no temporal relationships within signal sequences and are limited to weakly nonlinear signals at low data rates. Besides, ~\cite{10196036,zhao2022modeling} focus solely on a single link without modeling TXs in many links with crosstalk. Thus, it remains an open question to develop an efficient model for high-speed link TXs.

\subsubsection{Why Transformer?}
The Transformer proposed by Vaswani et al.~\cite{vaswani2017attention} excels in dealing with long-range dependencies and parallel computation with the encoder-decoder structure. To achieve non-sequential inputs of link parameters and sequential output signals, it is natural to pair the Transformer decoder with a non-sequential encoder.
Unlike RNNs or LSTMs, the Transformer decoder can manage non-sequential input through its attention mechanism instead of relying on strict sequential information unfolding over time, while effectively modeling highly nonlinear TX signals with long-range dependencies.

%% file: docs/prelim.tex
\section{Basics of Non-Autoregressive Transformer}
\subsection{Basic Structures of Transformer}
\subsubsection{Multi-Head Attention Mechanism.}
The attention mechanism is the key innovation of Transformer, capturing long-range dependencies by allowing each element to attend over other elements regardless of distance, enabling to process the entire sequence simultaneously~\cite{vaswani2017attention}. With the token embeddings transformed into query ($Q$), key ($K$) of dimension $d_k$, and value vectors ($V$) by multiplying with three learnable matrices $W^Q$, $W^K$ and $W^V$, the attention mechanism calculates scaled dot products of the query with all keys and applies softmax to obtain the weights for the values as:
\begin{equation}
\operatorname{Attention}(Q, K, V)=\operatorname{softmax}\left(\frac{Q K^T}{\sqrt{d_k}}\right) V,
\label{eq:attention}
\end{equation}
which allows integrating contextual information across the entire sequence by the summation of values weighted by query-key match.

Multi-head attention runs the above mechanism in parallel across independent attention heads. Each head $i$ has its own parameters, $W^Q_i$, $W^K_i$ and $W^V_i$. The outputs from each head are then concatenated and projected with $W^O$ as follows:
\begin{equation}
\begin{aligned}
\operatorname{MultiHead}(Q, K, V) & =\operatorname{Concat}\left(\operatorname{head}_1, \ldots, \operatorname{head}_{\mathrm{h}}\right) W^O, \\
\text { where } \operatorname{head}_i & =\operatorname{Attention}\left(Q W_i^Q, K W_i^K, V W_i^V\right).
\end{aligned}
\end{equation}
The combination of information from all heads provides a multi-perspective representation of the sequence, enhancing the model's ability to focus on various parts of the sequence concurrently and understand complex relationships.

\subsubsection{Positional Encoding.}
The order of a sequence decisively influences its connotation, with RNNs and LSTMs naturally recognizing this through sequential processing. The attention mechanism processes the token relevance regardless their position, potentially causing randomization and error. Positional encoding (PE) is applied to the token embeddings before the attention mechanism processes the sequence, which assigns each token a position-specific vector to maintain their order information using an embedding matrix~\cite{vaswani2017attention,gehring2017convolutional}:
\begin{equation}
\label{eq:PE}
\begin{aligned}
P E_{(p o s, 2 i)} & =\sin \left(p o s / 10000^{2 i / d_{\mathrm{model}}}\right), \\
P E_{(p o s, 2 i+1)} & =\cos \left(p o s / 10000^{2 i / d_{\mathrm{model}}}\right),
\end{aligned} 
\end{equation}
where $d_{\mathrm{model}}$ is the embedding dimension typically equal to the model dimension, $i$ is the index of the embedding dimension and $pos$ is the position within the sequence. By adding the positional embeddings to the token embeddings, Transformer provides the representation of each token with its position information included.

\subsubsection{Position-Wise Feed-Forward Networks.}
Position-wise Feed-Forward Networks (FFNs) consist of two dense layers mapping from $d_{\mathrm{model}}$ to another space of dimension $d_{ff}$ and then back, which is separately and identically applied to each position in the sequence~\cite{vaswani2017attention}. The FFN function with a ReLU activation is defined as:
\begin{equation}
\operatorname{FFN}(x)=\max \left(0, x W_1+b_1\right) W_2+b_2,
\end{equation}
where $W_1$, $W_2$, $b_1$, $b_2$ are learnable parameters.

\subsection{Non-Autoregressive Mechanism}

\begin{figure}[tb]
\centering
\includegraphics[width=\linewidth]{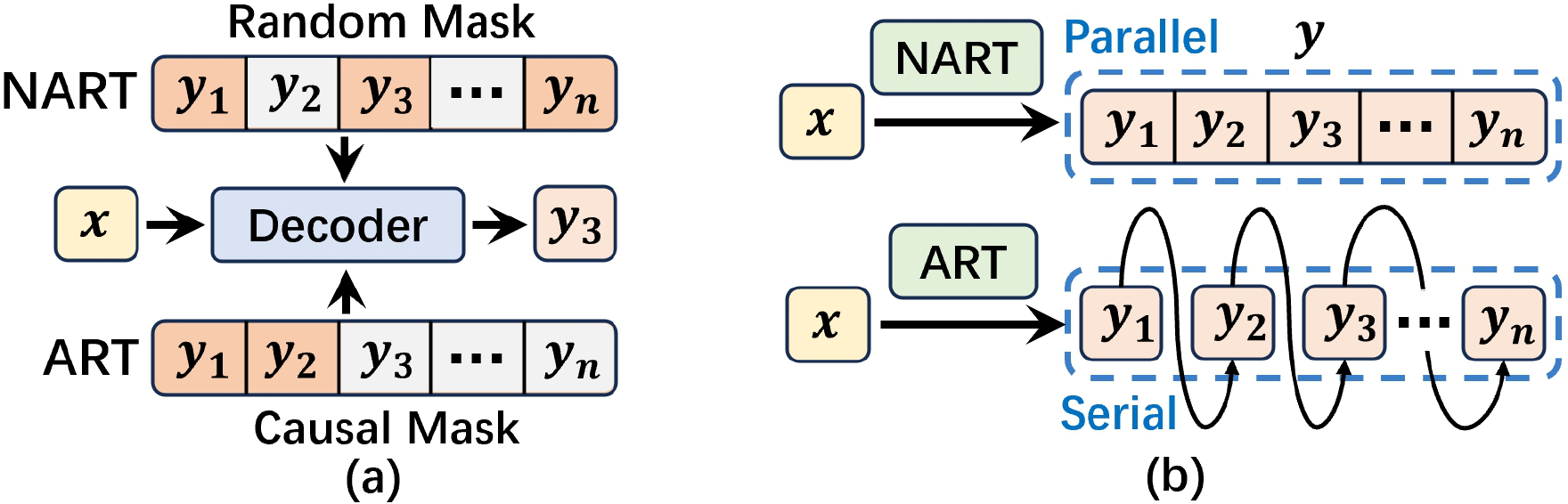}
\caption{Comparison of NART and ART: (a) training process; (b) inference process.}
\label{fig:AT-NAT}
\end{figure}

For sequence modeling tasks, most models adopt autoregressive (AR) methods, including AR Transformers (ARTs)~\cite{li2021long,zerveas2021transformer}. Given an input source $ \boldsymbol{X}$, ARTs generate a target sequence $\boldsymbol{y}={y_1, y_2, \ldots,y_n}$ of length $n$ through a chain of conditional probabilities:
\begin{equation}
p(\boldsymbol{y} \mid \boldsymbol{X})=\prod_{i=1}^n p\left(y_i \mid y_{<i}, \boldsymbol{X}\right),
\end{equation}
where $y_{<i}$ denotes the target tokens generated from $\boldsymbol{X}$ before the $i^{th}$ token. Conditioning each token on its previous ones, ARTs have to perform $n$ iterations to sequentially decode each $y_i$, making inference very time-consuming. To enhance the inference speed, NAR Transformers (NARTs) are gaining increasing attention for their efficient parallel decoding capabilities~\cite{lee2018deterministic,ghazvininejad2019mask, chen2020non, huang2022directed}. Assuming that output tokens are conditionally independent given $\boldsymbol{X}$, the NAR mechanism cancels out the left-to-right dependency and decomposes the conditional dependency chain of $p(\boldsymbol{y} \mid \boldsymbol{X})$ as:
\begin{equation}
p(\boldsymbol{y} \mid \boldsymbol{X})=\prod_{i=1}^n p\left(y_i \mid \boldsymbol{X}\right),
\end{equation}
which indicates that $y_i$ relies only on the source $\boldsymbol{X}$, allowing simultaneous sequence decoding. ~\Cref{fig:AT-NAT} (a) and (b) respectively compares the training and inference process between NARTs and ARTs. ARTs employ a causal mask to ensure predicting $y_i$ based only on prior tokens during training, and sequentially decode each $y_i$ during inference, while NARTs apply random masking to remove sequential dependencies and decodes the entire sequence $\boldsymbol{y}$ in parallel~\cite{ghazvininejad2019mask}.

%% file: docs/formulation.tex
\section{Problem Formulation}

\subsection{Formulation of TX Dynamics within Links}
\subsubsection{Overview.}
\begin{figure}[tb]
\centering
\includegraphics[width=\linewidth]{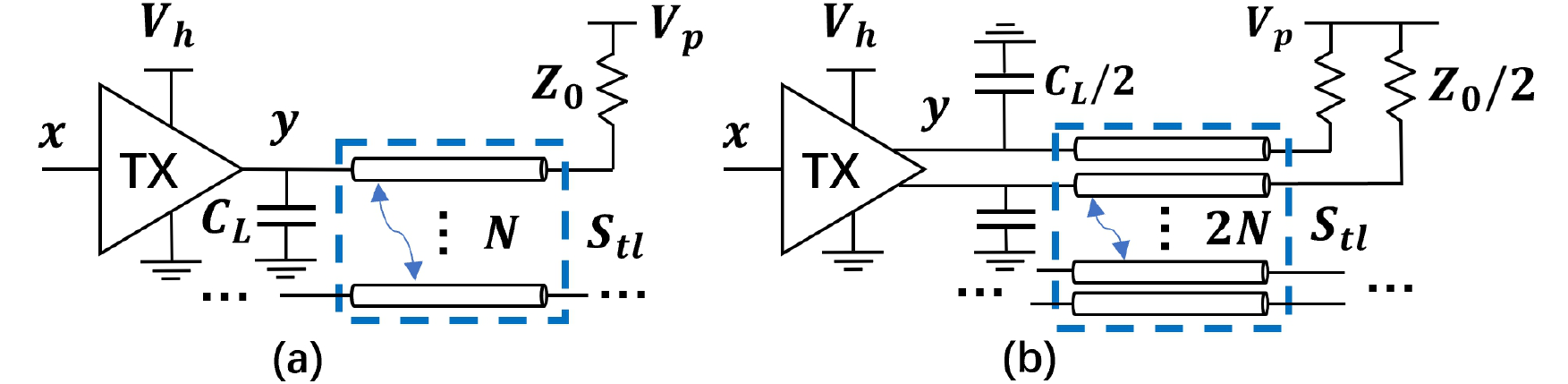}
\caption{Equivalent circuit model for TXs in an N-link system: (a) single-ended TXs; (b) differential TXs.}
\label{fig:eq-circuit}
\end{figure}
\Cref{fig:eq-circuit} depicts the equivalent circuit used for analyzing TXs in an N-link system: (a) for single-ended and (b) for differential TXs. Each TX connects to a load capacitance $C_L$, one (two) non-ideal transmission line(s) characterized by S-parameters $\boldsymbol{S}_{tl}$, and a pull-up impedance $Z_0$ modeling the RX design with a pull-up level $V_p$. 
Reference lines are omitted for simplicity. The input data stream ${\boldsymbol{x}}$ transmitted by each TX can be formatted as a series of $m$ binary symbols ${\boldsymbol{x}}_{s}= x_0, x_1, \ldots, x_{m-1}$ where $x_i = \text{``0'', ``1''}, i = 0, \ldots, m-1$. A two-tap finite impulse response (FIR) equalizer is typically located in the TX, which operates through a first-in first-out (FIFO) buffer to linearly adjust the transmitted voltage levels, attenuating low-frequency components and mitigating channel loss. The equalized data sequence ${\boldsymbol{x}}^d_{s}=x^d_0,x^d_1,\ldots,x^d_{m-1}$ of ${\boldsymbol{x}}_{s}$ is calculated by:
\begin{equation}
\begin{split}
      x^d_i = H_0x_i + H_1x_{i-1}, \qquad &i=1,\ldots,m-1 \\
    \text { where }\ |H_0|+|H_1| = 1, \qquad &H_0>0, H_1 < 0  
\end{split}
\end{equation}
where $H_0$ and $H_1$ are the taps of the filter. Due to parasitic coupling of transmission lines, each link potentially interferes with the TX of every other link through NEXT. We aim to predict the interfered TX output signals ${\boldsymbol{y}}$ in an N-link system given input signal sequences ${\boldsymbol{x}}$ and link parameters $H_0,V_h, C_L, \boldsymbol{S}_{tl}, Z_0, V_p$.

\subsubsection{Parameterization of Input Signal.}
To transmit ${\boldsymbol{x}}_{s}$ through the links, NRZ (Non-Return-to-Zeros) signaling uses trapezoidal wave with each level corresponding to ``0'' or ``1'', characterized by its amplitude $V_h$ matching the supply voltage, a signal period $t_p$ and transition time $t_{rf}$ assuming equal rising and falling phase. PAM4 uses equally spaced 4 distinct levels of $V_h$ to carry two bits (``00'', ``01'', ``10'', ``11'') per pulse, with a pulse period $t_p$ and transition time $t_{rf}$ between two levels, doubling the data rate. 
Despite $t_{rf}$'s drastic variations across signal periods, its proportion remains stable. We use the ratio $r_{rf}=\frac{t_{rf}}{t_p}$ to represent transition time.
Therefore, we could fully describe the input signal ${\boldsymbol{x}}$ by the symbol sequence ${\boldsymbol{x}}_{s}$ and its waveform parameters of $V_h$, $t_p$ and $r_{rf}$.

\subsubsection{Interfered Output Decomposition. }
Given that the intrinsic output of each TX is independent and is a foundation upon which crosstalk independently contributed by the other $N-1$ links is superimposed, the interfered output $\boldsymbol{y}_{i}$ of TX$_i$ could be decomposed into its intrinsic output $\boldsymbol{y}^{intr}_{i}$ absent any crosstalk plus the accumulated crosstalk from other $N-1$ links, expressed as:
\begin{equation}
    \boldsymbol{y}_{i}=\boldsymbol{y}^{intr}_{i}+\sum_{j=1, j \neq i}^{N} \boldsymbol{C}_{ij} ,
\end{equation}
where $\boldsymbol{C}_{ij}$ is the NEXT from link$_j$ onto link$_i$. For problem regularity, we evaluate the intrinsic output $\boldsymbol{y}^{intr}_{i}$ in a basic 2-link setup. Analyzing TX$_i$ in an $N$-link system could then be simplified into analyzing $N-1$ instances of 2-link systems of link$_i$ and link$_j$, focusing on only one input signal each time: (1) determine $\boldsymbol{y}^{intr}_{i}$ given link$_i$'s input $\boldsymbol{x}_{i}$ which is conducted once with an arbitrary link$_j,  j \neq i$, and (2) determine $\boldsymbol{C}_{ij}$ given link$_j$'s input $\boldsymbol{x}_{j}$,  which is repeated $N-1$ times for $j=1,\ldots,N, j \neq i$.

\subsection{Model Formulation}
\label{subsec:model-formulation}
As per analysis, our objective is summarized as characterizing TX$_i$'s intrinsic output  $\boldsymbol{y}^{intr}_{i}$ and the crosstalk component $\boldsymbol{C}_{ij}$ separately in a 2-link system. We introduce a boolean variable $K$ to differentiate these two cases: $K$=0 for $\boldsymbol{y}^{intr}_{i}$ and $K$=1 for $\boldsymbol{C}_{ij}$. The TX model could then be formulated by the function $f$ as:
\begin{equation}
\label{eq:formulation}
    \boldsymbol{y}_K = f(K, H_0, {\boldsymbol{x}}_{s}, V_h, t_p, r_{rf}, C_L, \boldsymbol{S}_{tl}, Z_0, V_p),
\end{equation}
where $\boldsymbol{S}_{tl}$ represents S-parameters of the transmission lines in a 2-link system, sized $4 \times 4$ for single-ended and $8 \times 8$ for differential TXs. With the interested TX$_i$'s link$_i$ always treated as the first link in the 2-link system, when $K=0$ or $1$, $\{{\boldsymbol{x}}_{s}, V_h, t_p, r_{rf}\}$ is considered the input signal parameters of the first link (victim) or the second link (aggressor), for intrinsic output or crosstalk prediction.

Despite voltage continuity, we reformulate the regression problem of~\Cref{eq:formulation} as a classification one to enhance model performance. 
We categorize the entire voltage range with the step size $\Delta v$ between each voltage category and construct a dictionary to map a voltage value to its nearest class. 
Since intrinsic and crosstalk components require different categorization granularity due to their magnitude mismatch, we use two dictionaries with equal length for model structural consistency - $\textbf{D}_I$ for intrinsic outputs ranging from $v^I_{\mathrm{min}}$ to $v^I_{\mathrm{max}}$ with $\Delta v^I$, and $\textbf{D}_C$ for crosstalk from $v^C_{\mathrm{min}}$ to $v^C_{\mathrm{max}}$ with $\Delta v^C$:
\begin{equation}
\textbf{D}_I = \left\{v_k : \textit{Class}_{k+\textit{1}} \mid v_k = v^I_{{min}} + k \cdot \Delta v^I, \ k = 0, 1, \ldots, \frac{v^I_{\mathrm{max}} - v^I_{\mathrm{min}}}{\Delta v^I} \right\},
\label{eq:D1}
\end{equation}
\begin{equation}
 \textbf{D}_C = \left\{v_k : \textit{Class}_{k+\textit{1}} \mid v_k = v^C_{\mathrm{min}} + k \cdot \Delta v^C, \ k = 0, 1, \ldots, \frac{v^C_{\mathrm{max}} - v^C_{\mathrm{min}}}{\Delta v^C} \right\}. 
 \label{eq:D2}
\end{equation}
$\textit{Class}_\textit{0}$ is assigned to a special \textit{<mask>} token in both $\textbf{D}_I$ and $\textbf{D}_C$. With sufficiently small $\Delta v$, we can accurately reconstruct the original voltage after categorizing it. Eq.~\ref{eq:formulation} is finally converted into:
\begin{equation}
    \boldsymbol{y}_c =f_c(K, H_0, {\boldsymbol{x}}_{s}, V_h, t_p, r_{rf}, C_L, \boldsymbol{S}_{tl}, Z_0, V_p)=f_c(\boldsymbol{X}),
\end{equation}
where $f_c$ is the function generating the voltage category sequence $\boldsymbol{y}_c$ from the input features $\boldsymbol{X}= \{K, H_0, {\boldsymbol{x}}_{s}, V_h, t_p, r_{rf}, C_L, \boldsymbol{S}_{tl}, Z_0, V_p\}$.

%% file: docs/model.tex
\section{Proposed Model Architecture}
\label{subsec:model-overview}

\begin{figure}[tb]
\centering
\includegraphics[width=\linewidth]{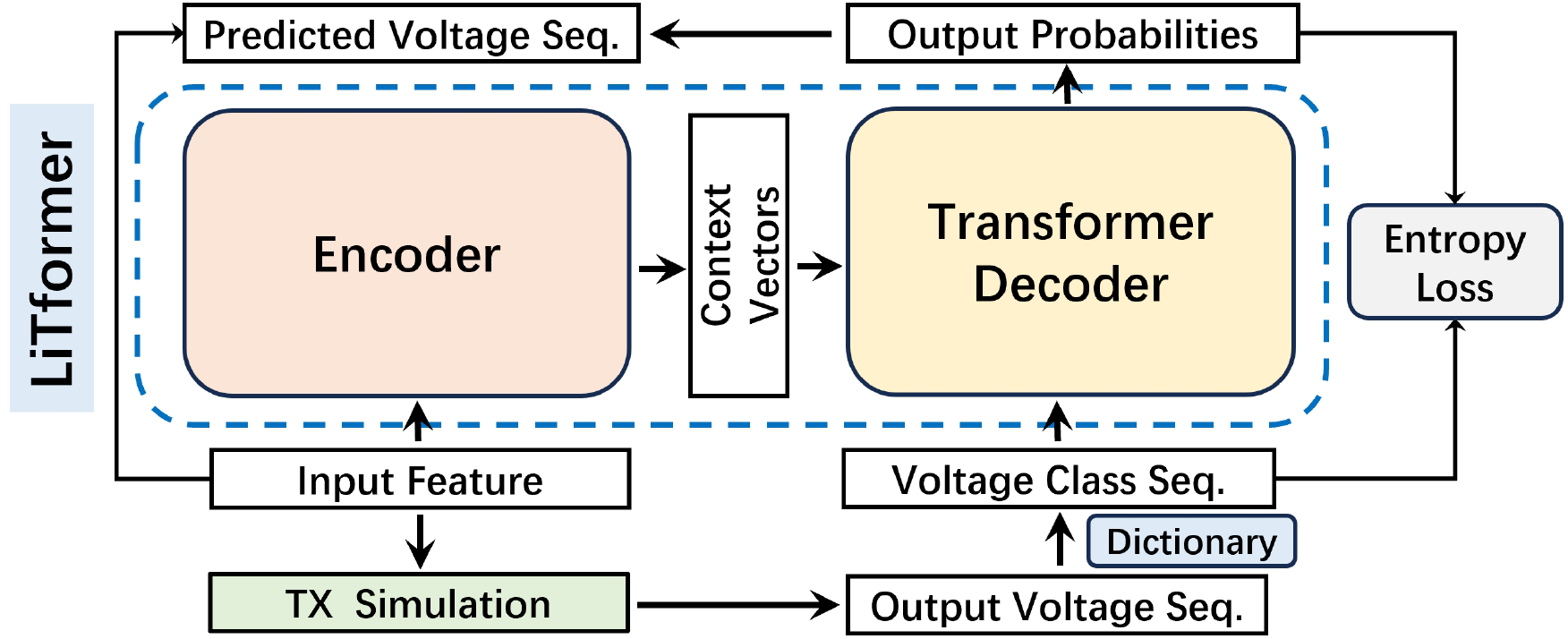}
\caption{Model Overview of the proposed LiTformer.}
\label{fig:model-overview}
\end{figure}

We implement an encoder-decoder architecture for \textit{LiTformer}, containing a non-sequential encoder and a sequential Transformer decoder as shown in~\Cref{fig:model-overview}. 
The non-sequential encoder processes unordered non-sequential inputs into context vectors with richer information. The decoder combines these vectors with the class sequence from the simulated TX outputs to generate token-wise class probability distributions, which is used for the cross-entropy (CE) loss calculation against the true class sequence for model optimization. 
These probability distributions transform into the output signal sequence during inference as detailed in~\Cref{subsec:decoding}. The proposed nonseq2seq architecture surpasses traditional seq2seq and static models in modeling nonlinear TXs with additional link parameters.

\subsection{Encoder Architecture}

\begin{figure}[t]
\centering
\includegraphics[width=1.05\linewidth]{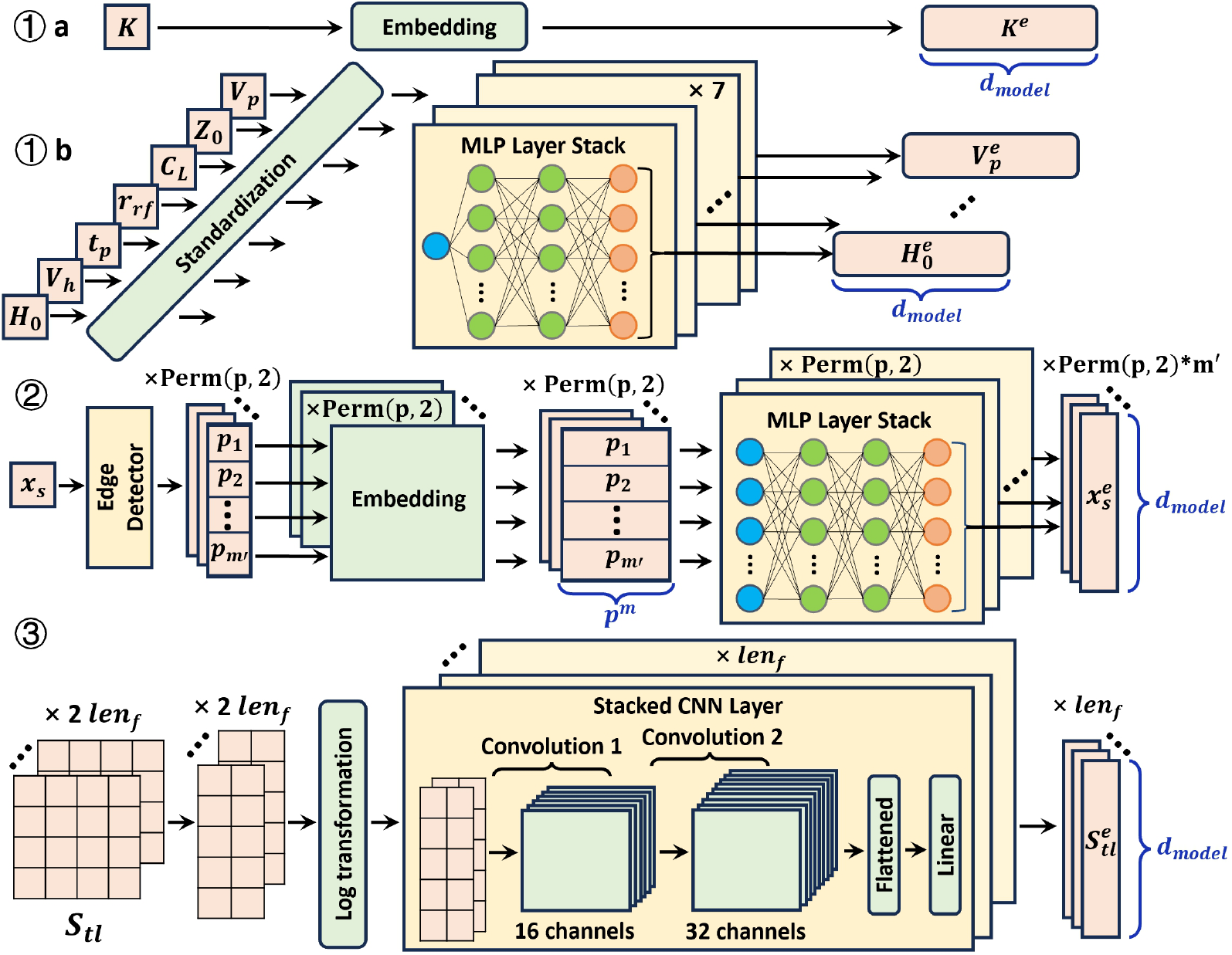}
\caption{Architecture of the LiTformer encoder: \textcircled{1} for scalar features; \textcircled{2} for the input signal sequence; \textcircled{3} for S-parameters.}
\label{fig:encoder}
\end{figure}

In this section, we introduce the various components of the proposed \textit{LiTformer} encoder as shown in~\Cref{fig:encoder}, which independently encodes the input features in~\Cref{subsec:model-formulation} into context vectors with a dimension of $d_\mathrm{model}$ for subsequent decoder processing.
\subsubsection{Scalar Features Encoding. }
\label{subsubsec:scalar Encoding}
We encode the binary variable $K$ as a $d_\mathrm{model}$-dimensional vector $K^e$ through a $2 \times d_\mathrm{model}$ embedding matrix as depicted in~\Cref{fig:encoder} \textcircled{1} a. For scalar features $H_0$, $V_h$, $t_p$, $r_{rf}$, $C_L$, $Z_0$, $V_p$, as shown in \textcircled{1} b, we standardize each by deducting the mean and dividing by the standard deviation to mitigate scale discrepancies and facilitate faster convergence. 
These standardized features are then converted into $d_\mathrm{model}$-dimensional vectors $H_0^e,\ldots, V_p^e$ via stacked Multi-Layer Perceptrons (MLP) layers — each feature fed into an individual single-input MLP with two hidden layers of 16 neurons, ReLU activation in the hidden layers, and linear output activation.

\subsubsection{Input Signal Sequence Encoding. }

To deserialize an $m$-symbol input sequence $\boldsymbol{x}_s$, we propose to detect its level transition positions for identification. For a signal modulated with $p$ levels, there are $\mathrm{Perm}(p, 2)$ possible directional transitions, where $\mathrm{Perm}(p, 2)$ represents the permutation number of arranging 2 out of $p$ distinct states, in total 2 types of edges for NRZ and 12 for PAM4. The transition from symbol $u$ to $v$ is denoted as $edge^{u\rightarrow v}$, where $u,v = 0,1,\ldots, p-1, u \neq v$. Each $x_i$ is assigned with a position index $i+1$ with valid positions from $1$ to $m$. $0$ is reserved to indicate an invalid position. 

Traversing the entire sequence of $\boldsymbol{x}_s$, we mark the position of $edge^{x_i\rightarrow x_{i+1}}$ at position $i+1$ for rising edges with $x_i<x_{i+1}$ or at $i$ for falling edges with $x_i>x_{i+1}$, where $i = 0,\ldots,m-2$. Since the signal rests at low before and after transmission, there is a rising $edge^{0\rightarrow x_0}$ at $1$ for nonzero $x_0$ and a falling $edge^{x_{m-1}\rightarrow 0}$ at $m$ for nonzero $x_{m-1}$. Each edge appears up to $m'$ times, where $m'=m/2$ for even $m$ or $m'=(m+1)/2$ for odd, with zeroes padding for uniformity. We summarize the edge detection algorithm in~\Cref{alg:edge_detector}, through which we achieve using non-sequential parameters to uniquely determine $\boldsymbol{x}_s$. For an NRZ ``1011'' sequence, $edge^{0\rightarrow 1}$ and $edge^{1\rightarrow 0}$ is located at $\{1, 3\}$ and $\{1,4\}$ respectively. For PAM4 ``0131'', $edge^{0\rightarrow 1}$, $edge^{1\rightarrow 3}$, $edge^{3\rightarrow 1}$ and $edge^{1\rightarrow 0}$ are located at $\{2, 0\}$, $\{3, 0\}$,$\{3, 0\}$ and $\{4, 0\}$, while all other edges are set to $\{0, 0\}$.

\begin{algorithm}[t]

\caption{Edge Detector for a p-Level m-Symbol Sequence}
\label{alg:edge_detector} 
\flushleft{\bf Input:}
A $p$-level m-symbol sequence $\boldsymbol{x}_s = x_0,x_1,\ldots,x_{m-1}$ \\
{\bf Output:}
Arrays of positions for each transition edge $edge^{u\rightarrow v}$ \\
\begin{algorithmic}[1]
\State \textbf{Set} $m' \gets m/2$ if $m$ is even, or $m' \gets (m+1)/2$ if $m$ is odd
\State \textbf{Initialize} An array for each $edge^{u,v}$, with $u,v = 0,1,\ldots, p-1, u \neq v$, of length $m'$ and filled with zeros
\State \textbf{Set} invalid position marker to $0$
\State Mark $edge^{0\rightarrow x_0}$ at position $1$ if $x_0 \neq 0$
\State Mark $edge^{x_{m-1}\rightarrow 0}$ at position $m$ if $x_{m-1} \neq 0$
\For{$i=1$ to $m-1$}
    \If{$x_{i-1} < x_i$} \State Mark $edge^{x_{i-1}\rightarrow x_i}$ at position $i+1$
    \ElsIf{$x_{i-1} > x_i$} \State Mark $edge^{x_{i-1}\rightarrow x_{i}}$ at position $i$
    \EndIf
\EndFor
\State \textbf{return} Arrays of positions for all $edge^{u\rightarrow v}$
\end{algorithmic}
\end{algorithm}

As shown in ~\Cref{fig:encoder} \textcircled{2}, $\boldsymbol{x}_s$ is transformed into $m'$ position indices for $\mathrm{Perm}(p,2)$ edges, which are embedded into a set of continuous $p^m$-dimensional vectors through $\mathrm{Perm}(p,2)$ instances of $(m+1) \times p^m$ embedding matrices unique to each $edge^{u\rightarrow v}$. These embeddings individually go through stacked MLPs as in~\Cref{subsubsec:scalar Encoding} adapted for $p^m$ input, with parameters unique to $edge^{u\rightarrow v}$. We finally obtain $\boldsymbol{x}^e_s$ containing $\mathrm{Perm(p, 2)} * m'$ vectors each of dimension $d_\mathrm{model}$. 

\subsubsection{S-parameter Encoding. }
We keep the frequency response nature of S-parameters, viewing their impact upon the system at each frequency as key factors. $\boldsymbol{S}_{tl}$ with $len_f$ frequency points can be decomposed into a real- and an imaginary-part matrix at each frequency. By exploiting their inherent symmetry to reduce redundancies, we streamline the model by considering only $\frac{n^2 + n}{2}$ effective values from an $n \times n$ matrix. We reshape the $4 \times 4$ ($8 \times 8$) $\boldsymbol{S}_{tl}$ with 10 (36) valid entries into $2 \times 5$ ($6 \times 6$) for 2-link single-ended (differential) TXs. Since S-parameters have quite small magnitude with a huge fluctuation in the order, and can be positive or negative, we shift their entries $s$ to positive and apply a logarithmic transformation to linearize and stabilize data distribution:
\begin{equation}
        s_{scaled} = log(s+1.1*min(\boldsymbol{S}_{tl})).
\end{equation}
The scaled matrices are processed through two consecutive Convolutional Neural Network (CNN) layers at each frequency point: the first CNN expanding the two input channels corresponding to the real and imaginary matrix into 16 outputs and the second further expanding into 32 outputs, both with a kernel of size 1 and ReLU activation. The resulting feature map is flattened and passed through a linear layer to produce a final $\boldsymbol{S}^e_{tl}$ with a dimension of $(len_f, d_\mathrm{model})$.
The $\boldsymbol{S}_{tl}$ encoding process is shown in ~\Cref{fig:encoder} \textcircled{3}.
 
The encoded features $\{K^e, H_0^e, V_h^e, t_p^e, r_{rf}^e, C_L^e, Z_0^e, V_p^e, \boldsymbol{x}_s^e, \boldsymbol{S}^e_{tl}\}$ are finally combined together into an unordered $d_\mathrm{model}$-dimensional vector $\boldsymbol{X}^e$ of length $8+len_f+\mathrm{Perm(p,2)}*m'$, which is subsequently fed into the decoder for interaction with the output sequence.

 \subsection{Decoder Architecture}

\begin{figure}[t]
\centering
\includegraphics[width=0.8\linewidth]{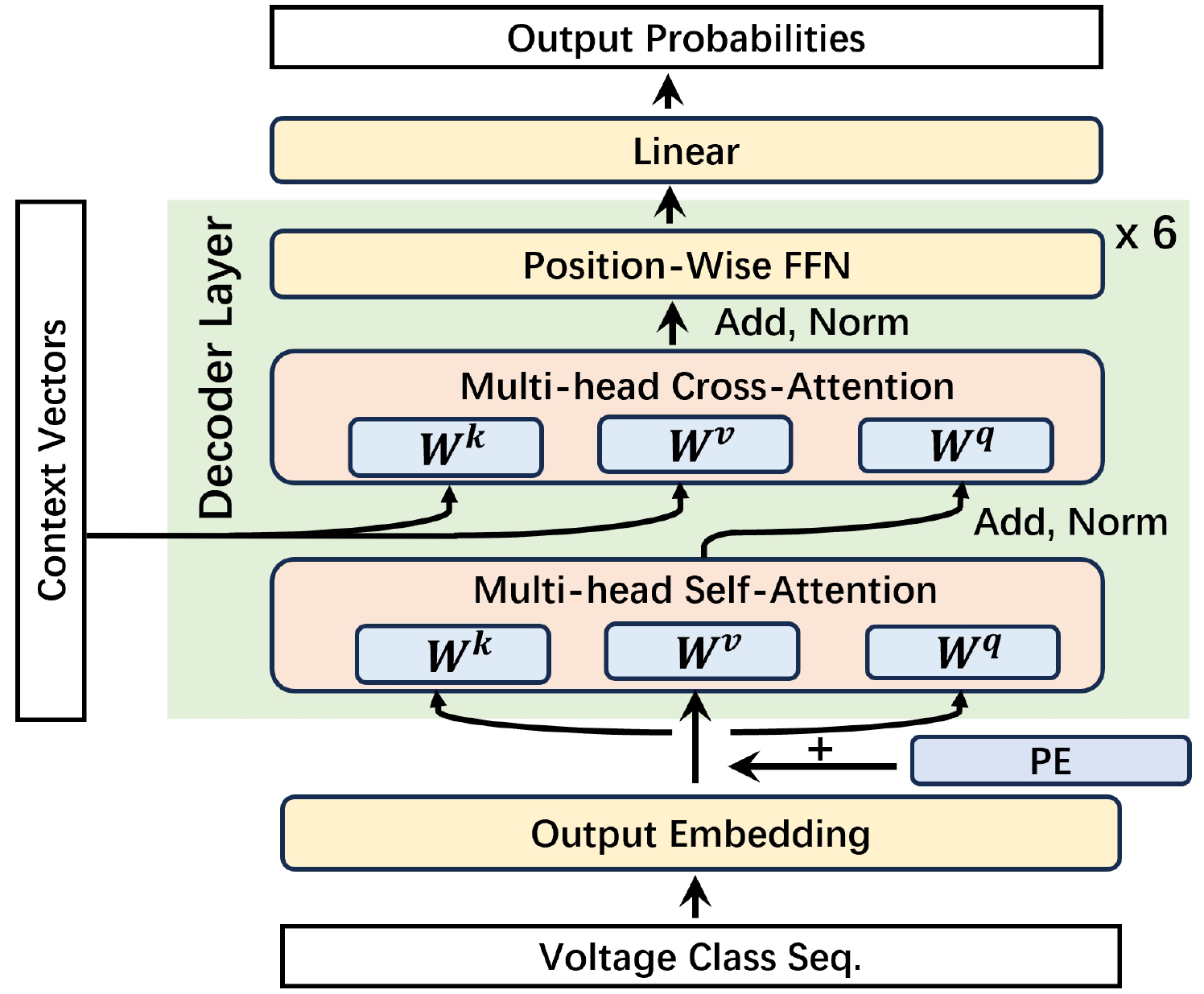}
\caption{Architecture of the LiTformer decoder.}
\label{fig:decoder}
\end{figure}

 We use a standard Transformer decoder architecture~\cite{vaswani2017attention} as shown in~\Cref{fig:decoder}. The voltage signal class sequence is embedded, added to the PE from Eq.~\ref{eq:PE}, and normalized for stability before entering the stacked 6 identical decoder layers. Each layer includes a multi-head self-attention module processing decoder outputs, a multi-head cross-attention module using context vectors for $K$ and $V$ and decoder outputs for $Q$ in Eq.~\ref{eq:attention}, which facilitates interaction between encoded non-sequential input features and the output sequence, followed by a position-wise FNN. The number of attention heads is all set to 8. Any two sub-layers are connected through residual connections and layer normalization. The final output is processed through a linear layer that translates dimensions 
 from $d_\mathrm{model}$ to dictionary size, determining output probabilities for each token in the output sequence.

%% file: docs/traintest.tex
\section{Training and Inference of NAR LiTformer}
\subsection{Randomly Masked Training}
To develop an NAR model, we modify the standard left-to-right decoder's attention mask to allow context integration from both sides for token prediction. Given the input feature $\boldsymbol{X}$ and a subset of target tokens $\boldsymbol{Z}$, the decoder calculates probabilities for a predetermined set of target tokens $\boldsymbol{Y}_{mask}$, treating them as conditionally independent. It predicts $P(y|\boldsymbol{X},\boldsymbol{Z})$ for each token $y$ in $\boldsymbol{Y}_{mask}$. A key advantage in the issue of modeling TX is that the output sequence length, denoted as $n$, is fixed based on the $m$-symbol input signal and unaffected by input variability. This eliminates the need for sequence length prediction, reducing the complexity associated with NAR models~\cite{ghazvininejad2019mask,lee2018deterministic} and boosting performance. 

In training, masked tokens are randomly selected as per~\cite{ghazvininejad2019mask}. We sample the number of masked tokens $n_{mask}$ from a uniform distribution between 1 and $n$, then randomly choose $n_{mask}$ tokens as $\boldsymbol{Y}_{mask}$, replacing their values with \textit{<mask>} which is designated $\textit{Class}_\textit{0}$ as in~\Cref{subsec:model-formulation}. The variety in masking schemes allows the model to adapt to scenarios ranging from easy (fewer masks) to challenging (more masks) when learning missing data~\cite{ghazvininejad2019mask}.
The decoder processes the masked sequence along with context vectors and generates an output sequence of the same length in one pass, instead of sequentially as in RNNs, using the attention mechanism that supports efficient parallel training. The model is trained using stochastic gradient descent to optimize the CE loss between predictions and target tokens across all $Y_{mask}$ tokens:
\begin{equation}
L_{CE}=-\sum_{y_i \in \boldsymbol{Y}_{mask}} \log P\left(y_i \mid \boldsymbol{X}, \boldsymbol{Z}\right).
\end{equation}
where $P\left(y_i \mid \boldsymbol{X}, \boldsymbol{Z}\right)$ represents the model's probability estimate for the true class of the masked token $y_i$. Although the decoder predicts the full sequence, the loss is calculated only on $\boldsymbol{Y}_{mask}$, directing the model to infer missing information.

\subsection{Inference with One-Pass Filtered Decoding}
\label{subsec:decoding}

\begin{figure}[t]
\centering
\includegraphics[width=\linewidth]{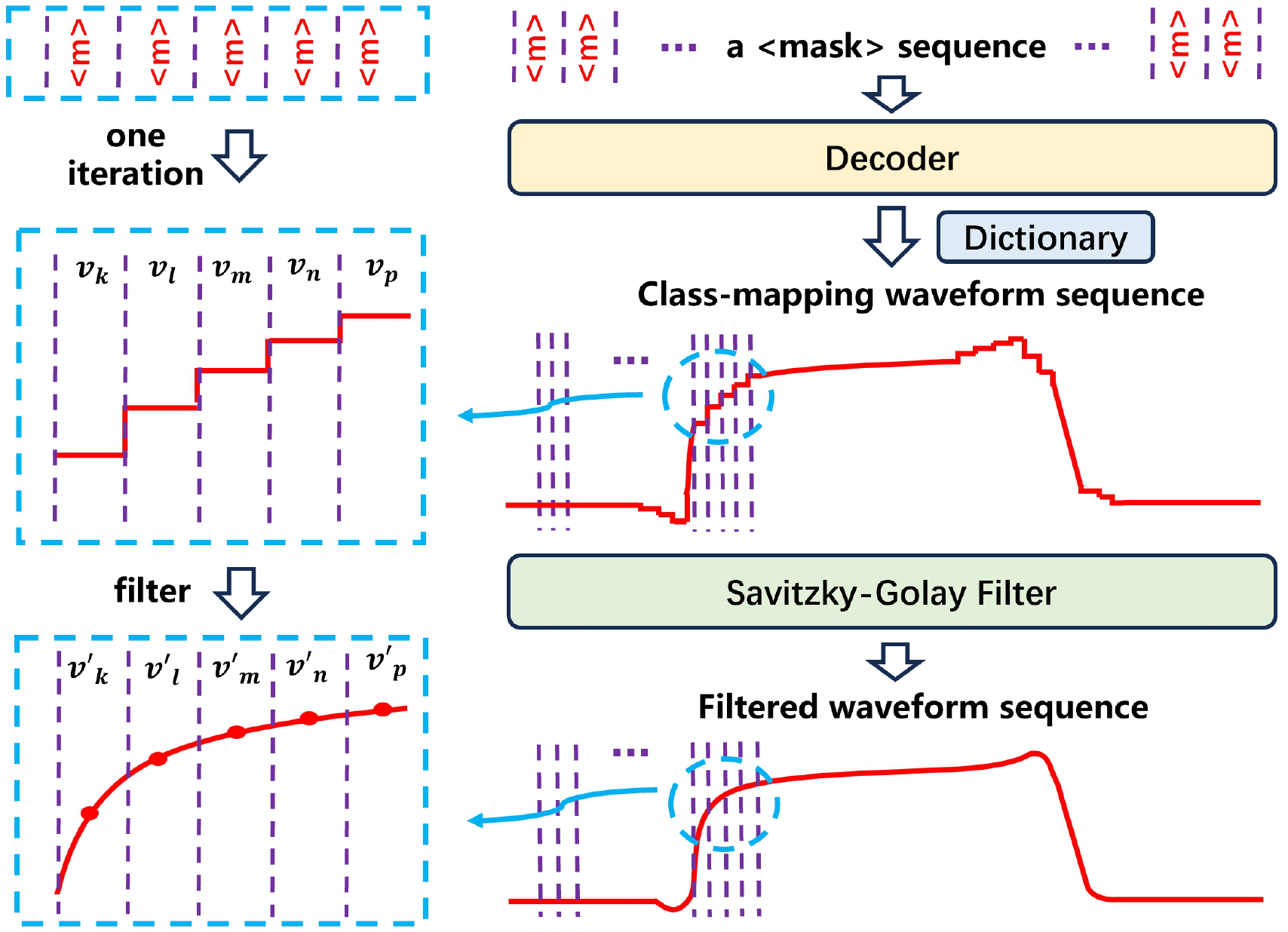}
\caption{Schematic of the proposed one-pass filtered decoding approach.}
\label{fig:1+1-decoding}
\end{figure}

After training, the model can simultaneously predict the masked tokens given $\boldsymbol{X}$. At the beginning of inference, without any information about the target, we input a fully masked sequence into the decoder and get all tokens predicted in parallel. We choose the class with the highest probability for each token:
\begin{equation}
y_c^* = \arg\max_{y_c \in \textbf{D}} P(y_c) ,
\end{equation}
where $\textbf{D}$ is the set of classes in $\textbf{D}_I$ or $\textbf{D}_C$. The output signal can then be reconstructed from $\textbf{D}_I$ or $\textbf{D}_C$. It is observed that the first-time decoded signal is already consistent with groundtruth, with minimal unevenness and deviation stemming from accuracy loss in categorization and model prediction. \textbf{One-time decoding is enough to achieve good performance.} We have also observed that \textbf{for signal waveform, iterative decoding essentially acts as a filter for decoding convergence,} which smooths signal irregularities without drastically altering its amplitude. Rather than iteratively refining uncertain predictions which risks introducing new errors without an explicit stopping condition~\cite{ghazvininejad2019mask}, we propose a one-pass filtered decoding approach utilizing waveform continuity. As shown in~\Cref{fig:1+1-decoding}, we apply a Savitzky-Golay filter~\cite{savitzky1964smoothing} on the first-time decoded and reconstructed signal sequence to smooth it and modify error to obtain the final output. With just a single parallel inference and a filtering process, we achieve efficient inference of the long-sequence signal without decoding latency, in contrast to AR methods which sequentially generate one output at a time. 

%% file: docs/exp.tex
\section{Experimental Results}
\subsection{Experiment Setup}
\begin{table}[tb]
\caption{Ranges of input signal parameters, link parameters except for $H_0$ between 0.8 and 1.0 and the line length.}
\label{tab:parameter-range}
\resizebox{\columnwidth}{!}{
\begin{tabular}{c|ccc|cccc}
\hline
\multirow{2}{*}{TX} & \multicolumn{3}{c|}{Input signal parameters} & \multicolumn{4}{c}{Link parameters} \\ \cline{2-8} 
 & $V_h$ (V) & $t_p$ (ps) & $r_{rf}$ (\%) & $C_L$ (pF) & $Z_0$ ($\Omega$) & $V_p$ (V) & $l$ (cm) \\ \hline
TS1 & 0.8$\sim$1.2 & 80$\sim$100 & 5$\sim$20 & 0.2$\sim$1.6 & 50$\sim$70 & 0.5$\sim$1.0 & 0.1$\sim$10 \\ \hline
TS2 & 0.8$\sim$1.2 & 100$\sim$150 & 5$\sim$20 & 0.01$\sim$0.5 & 40$\sim$70 & 0.4$\sim$0.8 & 0.1$\sim$10 \\ \hline
TD3 & 1.2$\sim$1.8 & 90$\sim$150 & 5$\sim$20 & 0.01$\sim$0.4 & 50$\sim$70 & 0.4$\sim$0.8 & 0.1$\sim$10 \\ \hline
TP4 & 0.8$\sim$1.5 & 60$\sim$150 & 10$\sim$20 & 0.05$\sim$0.5 & 50$\sim$70 & 0.6$\sim$1.0 & 0.5$\sim$10 \\ \hline
\end{tabular}
}
\end{table}

\begin{table}[tb]
\caption{Training and test errors of LiTformer and accuracy comparison of intrinsic and crosstalk outputs between the proposed LiTformer and SPICE.}
\label{tab:fund-results-1}
\resizebox{\columnwidth}{!}{
\begin{tabular}{c|c|c|cc|cc}
\hline
 &  &  & \multicolumn{2}{c|}{Intrinsic output} & \multicolumn{2}{c}{Crosstalk output} \\ \cline{4-7} 
\multirow{-2}{*}{TX} & \multirow{-2}{*}{\begin{tabular}[c]{@{}c@{}}Train\\ error\end{tabular}} & \multirow{-2}{*}{\begin{tabular}[c]{@{}c@{}}Test\\ error\end{tabular}} & CE & Mean AE/RE & CE & Mean AE/RE \\ \hline
TS1 & 0.335 & 1.97 & 2.64 & 4.75mV/0.61\% & 1.31 & 0.36mV/1.90\% \\ \hline
TS2 & 0.362 & 1.95 & 2.68 & 3.47mV/0.48\% & 1.22 & 0.29mV/2.19\% \\ \hline
TD3 & 0.350 & 2.74 & 3.71 & 7.36mV/1.26\% & 1.77 & 0.49mV/3.32\% \\ \hline
TP4 & 0.328 & 2.41 & 3.32 & 5.53mV/0.95\% & 1.49 & 0.39mV/1.93\%\\ \hline
\end{tabular}
}
\end{table}

\begin{table*}[]
\caption{Accuracy and runtime comparison for TX outputs in a 2-link and 16-link system between our LiTformer and SPICE.}
\label{tab:m-link}
\begin{tabular}{c|c|ccc|c|ccc}
\hline
\multirow{3}{*}{TX} & \multicolumn{4}{c|}{2-link system} & \multicolumn{4}{c}{16-link system} \\ \cline{2-9} 
 & Accuracy evaluation & \multicolumn{3}{c|}{Runtime comparison} & Accuracy evaluation & \multicolumn{3}{c}{Runtime comparison} \\ \cline{2-9} 
 & Mean AE/RE & LiTformer & SPICE & Speedup & Mean AE/RE & LiTformer & SPICE & Speedup \\ \hline
TS1 & 5.95mV/0.75\% & 6.51ms & 1.28s & 197x & 7.67mV/0.97\% & 20.3ms & 8.23s & 405x \\ \hline
TS2 & 5.55mV/0.78\% & 6.21ms & 2.41s & 388x & 4.98mV/0.68\% & 20.0ms & 13.1s & 655x \\ \hline
TD3 & 7.04mV/1.25\% & 9.21ms & 1.36s & 148x & 6.63mV/1.18\% & 39.4ms & 15.9s & 404x \\ \hline
TP4 & 5.75mV/0.96\% & 7.23ms & 3.30s & 456x & 6.50mV/1.07\% & 21.4ms & 20.2s & 944x \\ \hline
\end{tabular}
\end{table*}

\begin{figure}[tb]
\centering
\includegraphics[width=\linewidth]{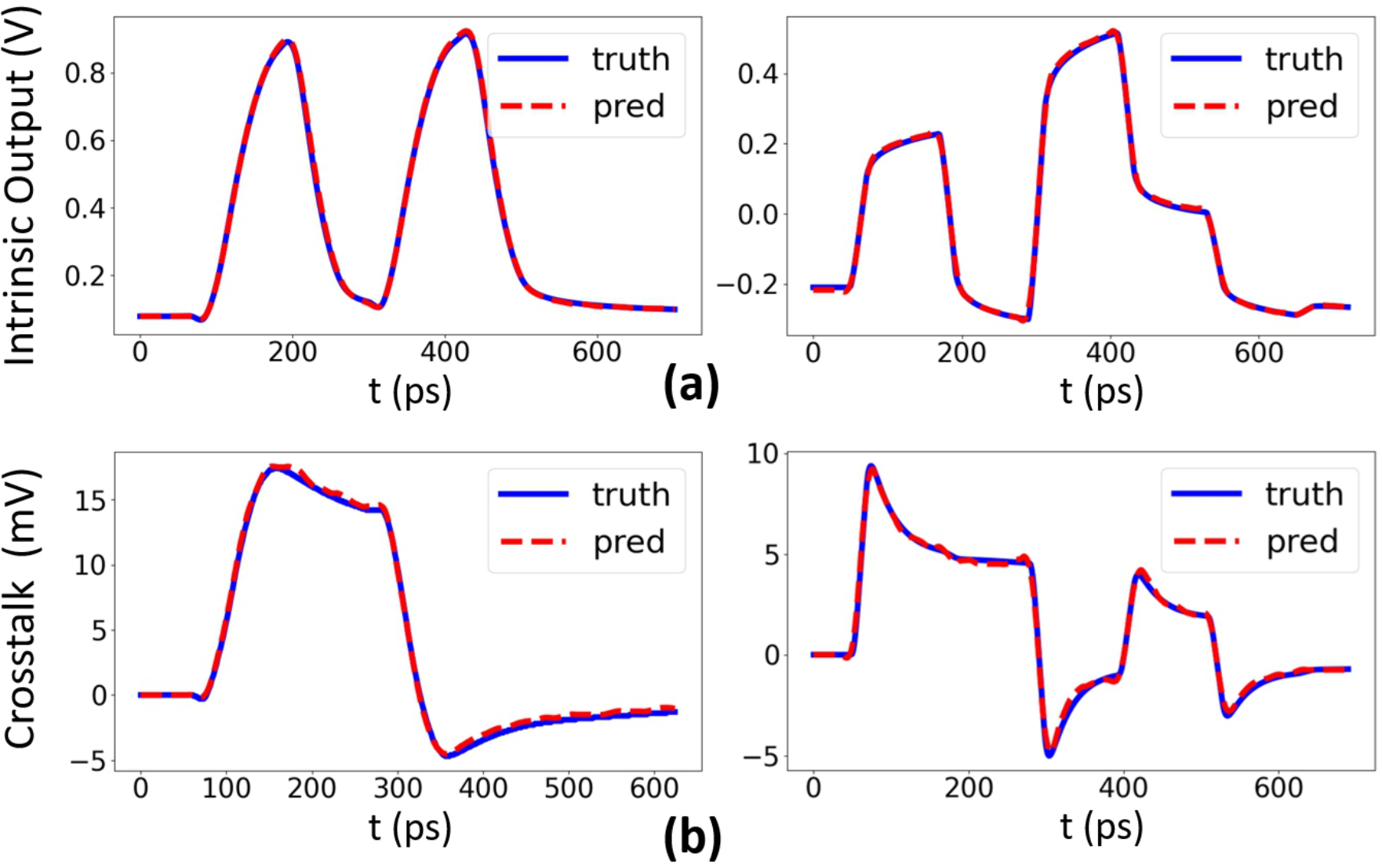}
\caption{Waveform comparison between the proposed LiTformer and SPICE for: (a) intrinsic output; (b) crosstalk component, on TS2 (left) and TP4 (right).}
\label{fig:crosstalk-intr}
\end{figure}

\begin{figure}[tb]
\centering
\includegraphics[width=\linewidth]{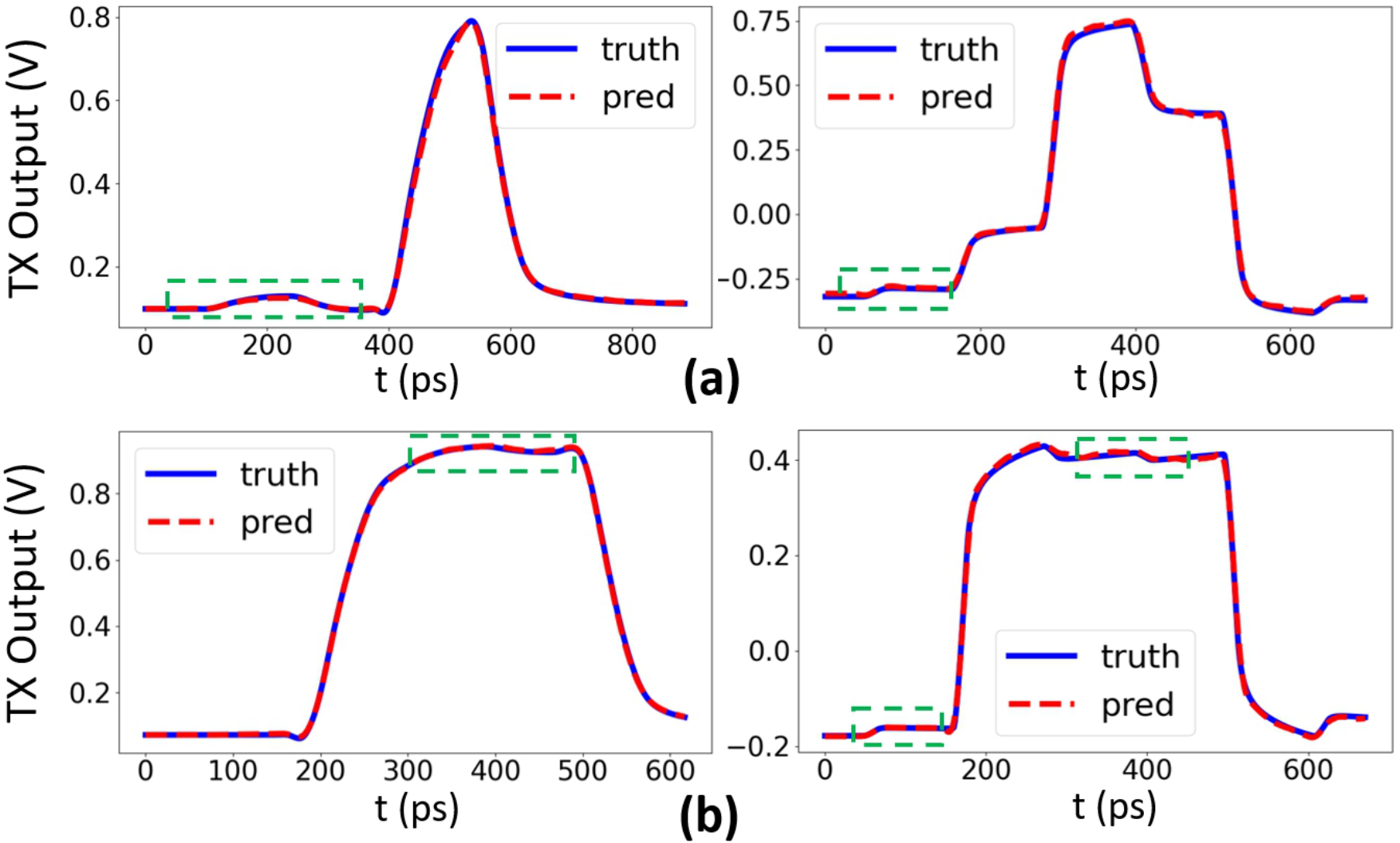}
\caption{Comparison between LiTformer and SPICE for TX interfered 
 waveforms in: (a) a 2-link system; (b) a 16-link system, on TS2 (left) and TP4 (right).}
\label{fig:2-16link}
\end{figure}

\begin{figure}[tb]
\centering
\includegraphics[width=\linewidth]{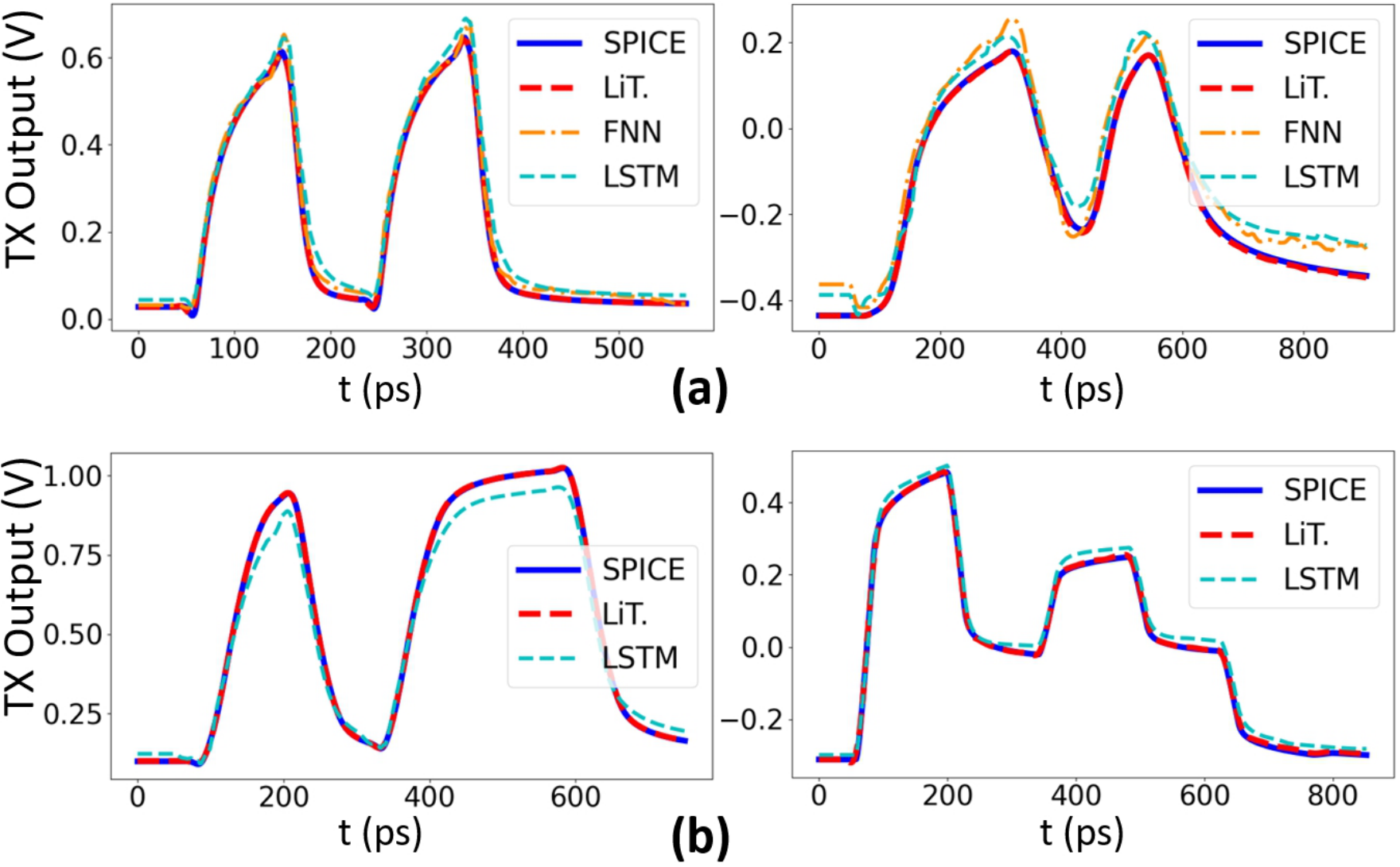}
\caption{Comparison of 1-link TX output between: (a) LiTformer, FNN~\cite{10196036}, LSTM~\cite{9395609} and SPICE with link parameters as model inputs on TS1 (left) and TD3 (right); (b)  LiTformer, LSTM~\cite{9395609} and SPICE with only the input signal as model inputs on TS2 (left) and TP4 (right).}
\label{fig:coutput-cmp}
\end{figure}

\begin{table*}[tb]
\tabcolsep=4pt
\caption{Accuracy and runtime comparison of 1-link TX outputs among: our LiTformer, FNN~\cite{10196036}, LSTM~\cite{9395609} and SPICE with link parameters as model inputs; our LiTformer, LSTM~\cite{9395609} and SPICE without link parameters as inputs.}
\label{tab:cmp-results}
\begin{tabular}{c|ccc|ccc|cc|cc}
\hline
\multirow{3}{*}{TX} & \multicolumn{6}{c|}{with link parameters} & \multicolumn{4}{c}{without link parameters} \\ \cline{2-11} 
 & \multicolumn{3}{c|}{Mean AE/RE  (mV/\%)} & \multicolumn{3}{c|}{Runtime Comparison} & \multicolumn{2}{c|}{Mean AE/RE (mV/\%)} & \multicolumn{2}{c}{Runtime Comparison} \\ \cline{2-11} 
 & \textbf{LiTformer} & FNN~\cite{10196036} & LSTM~\cite{9395609} & \textbf{LiTformer} & FNN~\cite{10196036} & LSTM~\cite{9395609} & \textbf{LiTformer} & LSTM~\cite{9395609} & \textbf{LiTformer} & LSTM~\cite{9395609} \\ \hline
TS1 & 4.20/0.53 & 21.0/2.78 & 26.0/3.50 & 6.89ms & 0.48ms & 9.37ms & 1.81/0.26 & 18.5/2.76 & 4.42ms & 8.65ms \\ \hline
TS2 & 2.81/ 0.38 & 30.4/4.11 & 33.9/4.64 & 6.00ms  & 0.38ms & 16.5ms & 0.86/0.11 & 38.2/4.98 & 4.78ms & 13.8ms \\ \hline
TD3 & 5.32/0.94 & 70.9/12.7 & 85.9/15.4 & 5.86ms & 0.26ms & 21.7ms & 1.36/0.25 & 67.9/12.2 & 4.64ms & 10.5ms \\ \hline
TP4 & 4.49/0.77 & 42.2/7.30 & 57.9/10.4 & 6.65ms & 0.31ms & 21.9ms & 3.28/0.56 & 28.0/5.04 & 4.97ms & 9.17ms  \\ \hline
\end{tabular}
\end{table*}

We evaluated our \textit{LiTformer} on 4 commercial high-speed TX designs: two NRZ single-ended TS1 and TS2, one NRZ differential TD3 and one PAM4 single-ended TP4. The dataset as groundtruth was derived from HSPICE transient simulation on a server with a 3.00GHz Intel Core i9-9980XE and 128GB RAM. We simulated 2-link TX outputs using random 4-symbol input sequences (4-bit for NRZ, 8-bit for PAM4) with uniformly sampled parameters as detailed in~\Cref{tab:parameter-range}, with $H_0$ between 0.8 and 1.0 for all cases. Due to the linear, periodic characteristics and band-limited response of transmission lines, we extracted $\boldsymbol{S}_{tl}$ using 5 frequency points per decade from 10Hz to 100GHz ($len_f = 51$) from different RLGC models of various lengths $l$. For intrinsic output samples $\boldsymbol{y}^{intr}_{i}$, we applied the input signal to the first link and kept the second low; for crosstalk $\boldsymbol{C}_{ij}$, we applied the input to the second with the first held high, and deducted $\boldsymbol{y}^{intr}_{i}$. We generated a 15k dataset with a balanced number of crosstalk and intrinsic samples to avoid bias and improve generalization, which was split into 12k for training, 1k for validation, and 2k for testing. LiTformer was trained in PyTorch on a GeForce RTX 4090 GPU.

We determined $\textbf{D}_I$ and $\textbf{D}_C$ by assessing the minimum and maximum values of intrinsic and crosstalk outputs for each TX dataset, to define the categorization range. $\textbf{D}_I$ spans a 1.6V range for TS1, TS2, TP4, and 1.8V for TD3, each with $\Delta v^I =  1mV$. $\textbf{D}_C$ ranges from -200 to 200mV with $\Delta v^C = 0.25mV$ for TS1, TS2, TP4, and -200 to 160mV with $\Delta v^C = 0.2mV$ for TD3. Including a \textit{<mask>}, dictionary lengths are 1602 for TS1, TS2, TP4 and 1802 for TD3. We set $d_{\text{model}}$ as 512 and trained the model with the Adam optimizer~\cite{kingma2014adam} with $\beta_1 = 0.9$, $\beta_2 = 0.98$, $\epsilon = 10^{-9}$, a learning rate of $0.0001$, and a batch size of $16$. To fully capture the 4-bit response of nonlinear TXs, we added a 1-bit tail for TS1, TS2, TP4, and a 3-bit tail for the more nonlinear TD3. The output sequence for all TXs was set to 501 points.

\subsection{Performance Evaluation}

\textit{LiTformer} was trained on the TS1, TS2, TD3, and TP4 datasets for 1280, 1160, 1240, and 1280 epochs respectively, taking about 16 hours. Training and testing CE losses are detailed in~\Cref{tab:fund-results-1} columns 2-3. LiTformer could accurately predict intrinsic outputs and crosstalk  considering various link parameters, with performance assessed by CE losses for class sequences and mean absolute and relative errors (AE/RE) of filtered reconstructed voltage signals shown in~\Cref{tab:fund-results-1} columns 4-7. RE is the AE ratio to the output amplitude. LiTformer achieves less than 1.26\% mean RE for intrinsic output prediction, establishing a robust foundation for overall TX output estimation. Though it reports crosstalk RE of 1.90-3.32\%, the corresponding AE is only around 0.3-0.5mV indicating minor amplitudes. Waveform comparisons with SPICE for TS2 and TP4 shown in~\Cref{fig:crosstalk-intr} reveal close matches for both intrinsic outputs and crosstalk.

To evaluate LiTformer's effectiveness of estimating TX interfered outputs in multiple links, we tested our LiTformer on 1000 simulated samples of 2-link and 16-link TX outputs (4 and 32 lines for TD3). We calculated the interfered output of TX$_i$ by adding the predicted intrinsic output with $K=0$ and the crosstalk pairing link$_i$ with each of the other links respectively with $K=1$. Accuracy and runtime comparisons are presented in~\Cref{tab:m-link}. Thanks to the accurate prediction of intrinsic outputs and crosstalk, LiTformer achieves mean REs of 0.75-1.25\% for 2-link TX outputs and 0.68-1.18\% for 16-link. By calculating intrinsic output and crosstalk components in one batch, LiTformer achieves inference time of 6-10ms with 148-456$\times$ speedup over SPICE for 2-link TXs and 20-40ms with 404-944$\times$ speedup for 16-link due to its parallel decoding capability, with SPICE's runtime growing drastically with more links. \Cref{fig:2-16link} illustrates the 2-link and 16-link TX output waveform comparisons to SPICE, with crosstalk components highlighted, indicating LiTformer's effectiveness in modeling TX dynamics in multiple links.

\subsection{Comparison with Related Work}

We evaluated our \textit{LiTformer} against FNN~\cite{10196036} and LSTM~\cite{9395609} for single-link TX modeling\footnote[1]{The FNN in \cite{10196036} and LSTM model in \cite{9395609} were not designed for multi-link TX modeling and hence single link is included for comparison.}. For single-link comparison, we removed LiTformer's encoder module of $K$ and input into CNNs the full S-parameters sized $2\times2$ for single-ended TXs, and S-parameters with effective elements sized $2\times5$ for TD3. The FNN took in the input signal sequence of a memory length, vector fitting parameters of $\boldsymbol{S}_{tl}$ and $H_0, C_L, Z_0, V_p$ to produce the output signal~\cite{10196036}. To incorporate link parameters, we combined the LSTM output~\cite{9395609} with them through an additional 4-layer FNN yielding the final output voltage. To specifically evaluate LiTformer's strength in modeling sequences with long-range dependencies, we also compared a further simplified LiTformer to LSTM~\cite{9395609} which both focused solely on input-output signal relationships without any link parameters. 

Accuracy and runtime comparisons for the two cases are detailed in~\Cref{tab:cmp-results}. Despite 10-hour training time, our LiTformer demonstrates significantly higher accuracy over FNN~\cite{10196036} and LSTM~\cite{9395609}, benefiting from its ability to effectively handle non-sequential parameters and capture long-range dependencies of the output sequence. The parallel decoding of LiTFormer also allows for a speed advantage over the recurrent LSTM~\cite{9395609}. Though both FNN~\cite{10196036} and LSTM~\cite{9395609} employ complex architectures to capture nonlinearities — with the FNN~\cite{10196036} having 37-40 hidden units and LSTM~\cite{9395609} up to 38 blocks and both input memory lengths set to around 500 — their performance on highly nonlinear signals at Gbps is still not optimal, even after extensive training. \Cref{fig:coutput-cmp} (a) and (b) visually compare waveforms in the two cases, underscoring our LiTformer's advantage over FNN~\cite{10196036} and LSTM~\cite{9395609} in modeling high-speed link TXs with high nonlinearity.

\section{Conclusions}
This work introduces LiTformer, an innovative Transformer-based model for high-speed link TXs, with a non-sequential encoder and a Transformer decoder accounting for link effects including crosstalk and long-range dependencies of signals. We employ an NAR mechanism in training and inference for fast prediction. Experiments show that in comparison to SPICE, the proposed LiTformer achieves a prominent speed with 148-456$\times$ speedup for 2-link TXs and 404-944$\times$ speedup for 16-link TXs while ensuring minimal error margins of 0.68-1.25\%, supporting 4-bit signals at Gbps data rates.

\begin{acks}
This work was supported in part by NSFC (Grant No. 62141404, and 62034007) and Major Program of National Natural Science Foundation of Zhejiang Province of China (Grant No. D24F040002).
\end{acks}